\documentclass{elsart}
\usepackage{longtable}
\usepackage{epsfig}
\begin{document}

\begin{frontmatter}

\title{Fast-neutron induced pre-equilibrium reactions on $^{55}$Mn 
and $^{63,65}$Cu at energies up to 40 MeV}
                                
\author[a]{M. Avrigeanu},
\author[b]{S.V. Chuvaev}, 
\author[b]{A.A. Filatenkov}, 
\author[c]{R.A. Forrest}, 
\author[d]{M. Herman}, 
\author[e]{A.J. Koning},
\author[f]{A.J.M.\ Plompen},
\author[a]{F.L. Roman},
\author[a]{V. Avrigeanu\corauthref{ca}}\ead{vavrig@ifin.nipne.ro}

\address[a]{"Horia Hulubei" National Institute for Physics and Nuclear 
	Engineering, P.O. Box MG-6, 077125 Bucharest, Romania}                
\address[b]{V.G. Khlopin Radium Institute, 2nd Murinski Ave. 28, 
	St. Petersburg 194021, Russia}
\address[c]{EURATOM-UKAEA Fusion Association, Culham Science Centre,
	Abingdon OX14 3DB, United Kingdom}
\address[d]{Brookhaven National Laboratory, P.O. Box 5000, Upton, 
	NY 11973-5000}
\address[e]{Nuclear Research and Consultancy Group NRG, P.O. Box 25, 
	1755 ZG Petten, The Netherlands}
\address[f] {European Commission, Joint Research Center, Institute for 
	Reference Materials and Measurements, B-2440 Geel, Belgium}
\corauth[ca]{Corresponding author. Tel.: +40-21-4046125}                                

\sloppy
\begin{abstract}
Excitation functions were measured for the 
$^{55}$Mn(n,2n)$^{54}$Mn, 
$^{55}$Mn(n,$\alpha$)$^{52}$V, 
$^{63}$Cu(n,$\alpha$)$^{60}$Co, 
$^{65}$Cu(n,2n)$^{64}$Cu, and
$^{65}$Cu(n,p)$^{65}$Ni reactions  from 13.47 to 14.83 MeV. The 
experimental cross sections are compared with the results of 
calculations including all activation channels for the stable 
isotopes of Mn and Cu, for neutron incident energies up to 50 MeV. 
Within the energy range up to 20 MeV the model calculations are most 
sensitive to the parameters related to nuclei in the early stages of 
the reaction, while the model assumptions are better established by 
analysis of the data in the energy range 20--40 MeV. While the present 
analysis has taken advantage of both a new set of accurate measured 
cross sections around 14 MeV and the larger data basis fortunately 
available between 20 and 40 MeV for the Mn and Cu isotopes, the need 
of additional measurements below as well as above 40 MeV is pointed out.
\end{abstract}

\begin{keyword}
$^{55}$Mn, $^{63,65}$Cu, E$\leq$40 MeV, Neutron activation cross 
section measurements, Nuclear reactions, Model calculations, Manganese,
Copper
\PACS 24.10.-I \sep 24.60.Dr \sep 25.40.-h \sep 28.20.-v
\end{keyword}
\end{frontmatter}

\section{Introduction}

As part of a general investigation  \cite{pr01,vs04,pr05,va06}
of the reaction mechanisms of fast neutrons at low and medium 
energies, we have analyzed the activation cross  sections of the
odd-mass isotopes $^{55}$Mn and $^{63,65}$Cu in the excitation-energy 
range up to 50 MeV. 

The main purpose of this paper is to present new experimental 
results and discuss some of the question marks associated with the
model calculations which combine pre-equilibrium emission (PE) with
equilibrium decay of the remaining compound nucleus. Although our
primary aim was to comply with the needs of a sound, complete and 
reliable neutron-induced cross section data library to address
safety and environmental issues of the fusion programme 
\cite{raf06,raf07}, the analysis results also enabled a stringent
test of models for the above-mentioned nuclear processes. The 
odd-mass target nuclei within the present work may be particularly
useful in connection with the proven influence of the $f_{7/2}$ 
neutron and proton shell closures on the particle PE spectra (e.g.,
\cite{mb85}). Actually Koning and Duijvestijn \cite{ajk04} pointed
out that omission of the shell effects is probably the most
important cause of the remaining discrepancies in their large-scale
comparison of the nucleon PE model with angle-integrated nucleon
spectra. Moreover, a systematic analysis by Mills et al. \cite{sjm91}
in the same mass range highlighted that some of the discrepancies 
observed in the yields of nuclides with closed or nearly-closed 
nucleon shells may not affect the inherent validity of the relevant 
model but follow the use of incorrect, e.g., average model parameters 
for certain nuclei involved in the decay process. Thus, in order
to gain insight into this problem, we have analyzed the activation
cross sections of $^{55}$Mn and $^{63,65}$Cu isotopes using the 
parameter databases obtained previously by global  optimization within 
the computer codes TALYS \cite{ajk04b} and EMPIRE-II \cite{mh04},
as well as a local parameter set within the STAPRE-H code \cite{ma95}. 
No fine tuning was done to optimize the description of the nucleon 
emission for all the cases, but for STAPRE-H a consistent set of local 
parameters has previously been established or validated on the basis 
of independent experimental information of, e.g., neutron total cross 
sections, proton reaction cross sections, low-lying level  and  
resonance data, and $\gamma$-ray strength functions  based  on  
neutron-capture data. The comparison of various calculations, 
including  their sensitivity to model approaches and parameters, has 
concerned all the activation channels for which there are measured 
data. It has thus avoided the use of model parameters which have 
been improperly adjusted to take into account properties peculiar to 
specific nuclei in the decay cascade, considered to be the case for 
discrepancies observed around the closures of both the $f_{7/2}$ 
proton and neutron shells \cite{sjm91}.

The cross sections for nuclear reactions induced by fast neutrons 
below 20 MeV are generally considered to be reasonably well known 
in spite of many fast neutron reactions for which the available data 
are either conflicting or incomplete even around 14 MeV. Consequently
the recent set of accurately measured cross sections below 20 MeV 
presented in Section 2 are extremely valuable. Actually the model 
calculations of these data are most sensitive to the parameters 
related to nuclei in the early stages of the reaction, including the 
PE processes which then dominate at higher energies. The corresponding 
model assumptions are thus better investigated by analysing the data 
above 20--30 MeV, which helpfully exist for the stable isotopes of Mn 
and Cu, provided that (i) a large body of data is described with no 
free but consistent model parameters, properly established by the 
analysis of other independent data, and (ii) the statistical parameters 
related to nuclei in the decay cascade are validated by the account 
of the data below $\sim$20 MeV. Thus, in order to match these former 
constraints, the main themes and parameters of the model calculations 
with the three computer codes are discussed in Section 3, while the 
experimental data and their comparison with the above-mentioned 
approaches are discussed in detail in Section 4. Finally, consideration 
of the main outcomes of this work is given in Section 5, the first 
half being additionally related to the above-mentioned analysis below 
$\sim$20 MeV, thus making possible the focus on the discussion of 
model assumptions at higher energies. 
Preliminary  results have been reported elsewhere \cite{aaf99,aaf03}.

\section{Experimental method}\label{sec:expmeth}

The well known activation method was used in order to obtain the
measured cross sections. A comprehensive overview of both the
measurements carried out at the KRI Neutron Generator NG-400 and
the experimental setup that was well tested in many previous 
measurements is given in Refs.\ \cite{aaf99,aaf01}, while we 
mention in the following only some basic points relevant to the 
present work.
       
Cross sections were determined by measuring the activity of samples
irradiated by neutrons from the $^3$H(d,n)$^4$He reaction, with a 
deuteron beam energy of 280 keV. Samples were made of tablets of 
pressed powder of potassium manganese KMnO$_4$, and respectively thin 
metallic copper foils of natural abundance. They had 14 mm diameter 
and the weight of 1100~mg and 360~mg, respectively. 
The target discs were sandwiched between two niobium or two aluminum 
foils that were used for neutron fluence determination. Sample 
assemblies were located around the target at different angles to the 
deuteron beam, thus providing different mean neutron energies in the 
region of 13.4 - 14.9~MeV. The neutron energy spectrum was calculated 
for every sample by taking into account the real geometry of each 
irradiation, the reaction cross section evaluation of Drosg \cite{md00}, 
and the stopping power evaluation of Anderson and Ziegler \cite{hha77}. 
The real beam and target characteristics were also accounted for 
\cite{aaf97}, and variations of the neutron flux during irradiation 
were registered by two independent scintillation detectors. The
absolute neutron fluences accumulated by the samples were determined 
by using the $^{93}$Nb(n,2n)$^{92m}$Nb and $^{27}$Al(n,$\alpha$)$^{24}$Na 
standard cross sections. The $^{27}$Al(n,$\alpha$)$^{24}$Na cross 
sections were taken from the FENDL/A-2.0 evaluation \cite{abp97}, 
while the experimental values of the $^{93}$Nb(n,2n)$^{92m}$Nb cross 
sections, obtained in Ref.\ \cite{aaf99} relative to the same standard,
were considered since these data show a smoother behavior of the cross 
section curve. Differences between the data of Ref.\ \cite{aaf99} and 
the evaluation of Ref.\ \cite{mw90} are less than the combined errors 
of measurement and evaluation, and do not exceed 1.35\%.
       
The $\gamma$-ray counting of the irradiated samples was done by means 
of two detectors, enclosed in passive shields. The first was a HPGe 
detector with a thin beryllium entrance window, and the second was a 
Ge(Li) detector. The HPGe detector had a relative efficiency of 24.7\%, 
and the Ge(Li) detector had a volume of 160 cm$^3$. The energy 
resolutions of the HPGe and Ge(Li) detectors were 1.8~keV and 4.0~keV, 
respectively, at 1332~keV. The background count rate was 
0.00064~counts/(s.keV) for the HPGe detector and 0.00019~counts/(s.keV) 
for the Ge(Li) detector at 1300~keV.
All the observable $\gamma$-ray peaks were revealed and identified 
in the spectra. The decay data used for cross section calculations 
(half-lives, $\gamma$-ray energies, and yields) were obtained from 
Ref. \cite{rbf95}. The reaction cross sections presented in 
Table \ref{tab:xsec} are weighted averages of cross sections 
obtained for every $\gamma$-ray line related to the reactions.

\section{Nuclear models and calculations}

\subsection{Global approach}

The two  sets  of global calculations within the direct-reaction, 
PE  and statistical Hauser-Feshbach (HF) models, performed by 
means of the computer codes TALYS \cite{ajk04b} and EMPIRE-II 
\cite{mh04}, have mainly used systematics based on global 
phenomenological analysis. Thus their results are firstly 
predictions of the reaction cross sections which should be
considered  from  the point  of  view  of  the  global parameters 
involved in the corresponding calculations. Actually, such blind
calculations   typically  produce  a  correct  shape   for   the
excitation  functions, while there is as much underprediction
as  overprediction when the results are compared with  data  for
all  nuclides  of  the periodic table of elements. Moreover, for  
a  true evaluation, a normalization of the curves can always be 
performed with nuclear model parameters that have an intrinsic
uncertainty,  such  as average radiative widths,  level  density
parameters  and  pre-equilibrium  matrix  elements. However, for 
large-scale data evaluations based on nuclear model calculations,
the performance of the corresponding global estimations of these 
data are also quite important. 
The main assumptions and parameters involved in this work 
for both sets of global calculations have been recently 
described \cite{va06}, while detailed descriptions were lately 
given \cite{ajk04,mh04} too. Therefore we give here only some 
specific points which have arisen in the meantime.

A similar approach using the code TALYS was applied in this work to 
the Mn and Cu isotopes as that reported in  Refs. \cite{vs04,pr05} 
for Co, Ni, and Mo  isotopes. However, a new version (0.72) of the 
TALYS code was used although the previous description regarding the 
choices for the optical model potential (OMP) \cite{ajk03}, the 
direct reaction (using  ECIS97 \cite{jr94}), the level density model 
\cite{ripl2,avi75} including the damping of  shell effects at high 
excitation energies, and especially the PE contributions with the 
two-component Exciton model using Kalbach systematics \cite{ck86} 
and particle-hole state densities including surface  effects 
\cite{ck00,ck85} which  depend on the type of projectile and the 
target mass \cite{ajk04} has been not altered for calculations that 
were performed in this work. The discrete level schemes are adopted 
from the RIPL-2 database \cite{ripl2}. We note that for this paper 
TALYS was only used for the global approach. A full description of 
the models and methods used in TALYS can be found in Refs. 
\cite{ajk06,mcd06,ajk07,ajk07b}, where also the applicability of 
the code for the local approach, i.e. optimized parameters for each 
nucleus, is demonstrated. On the other hand, the 2.19 version of 
the nuclear reaction code EMPIRE-II has been used for this work due
to its advantage of including the PE exciton model for cluster 
emission \cite{mh04}. At the same time, besides the adoption of
default parameters, the Hybrid Monte-Carlo simulation approach 
has been selected for the nucleon PE due to our interest 
in the neutron energies higher than 30 MeV.

\subsection{Local approach}

The particular properties of various target nuclei and reaction 
channels have been considered by using a consistent local parameter  
set, established on the basis of various independent data in a small 
range of mass and charge numbers. A generalized Geometry-Dependent 
Hybrid (GDH) model \cite{mb83,mb73} for PE processes in STAPRE-H 
version of the original code \cite{mu81} includes the 
angular-momentum conservation \cite{ma88} and the $\alpha$-particle 
and deuteron emission based on a pre-formation probability $\varphi$ 
\cite{eg81} with the values in the present work of 0.2 for 
$\alpha$-particles and 0.4 for deuterons \cite{vs04}. The same optical 
potential and nuclear level density parameters have been used in the 
framework of the OM \cite{ob81}, GDH and HF models, for calculation 
of the intra-nuclear transition rates and single-particle level (s.p.l.)
densities at the Fermi level \cite{mb73,ma94,ma98}, respectively, in 
the former case. 

The nucleon optical potential of Koning and Delaroche \cite{ajk03},
used by default in both TALYS and EMPIRE codes, has obviously been 
the first option. However, a basic point revealed by these authors 
is that their global potential does not reproduce the minimum around 
the neutron energy of 1-2 MeV for the total neutron cross sections 
of the $A$$\sim$60 nuclei. Following also their comment on the 
constant geometry parameters which may be responsible for this aspect, 
we have applied the SPRT method \cite{jpd76} for determination of the 
OMP parameters over a wide neutron energy range through analysis of 
the $s$- and $p$-wave neutron strength functions, the potential 
scattering radius $R'$ and the energy dependence of the total cross 
section $\sigma_T (E)$. The recent RIPL-2 recommendations \cite{ripl2} 
for the low-energy neutron scattering properties and the available 
measured $\sigma_T$ data (Fig. 1) have been used in this respect, and
we found that it is necessary to consider the energy dependence of the 
real potential geometry at lower energies shown in Table \ref{tab:nres}. 
These potentials were used also for the calculation of the collective 
inelastic scattering cross sections by means of the direct-interaction 
distorted-wave Born approximation (DWBA) method and a local version 
of the computer code DWUCK4 \cite{pdk84}. The weak coupling model was 
adopted in this respect for the odd nuclei $^{55}$Mn and $^{63,65}$Cu 
using the collective state parameters of Kalbach \cite{ck00}. Typical 
ratios of the direct inelastic scattering to the total reaction cross 
sections in the energy range from few to 60 MeV decrease from $\sim$11 
to 5\%, for the $^{55}$Mn nucleus, and from $\sim$8 to 3\% for the Cu 
isotopes.

The OMP of Koning and Delaroche \cite{ajk03} was considered also for the 
calculation of proton transmission coefficients on the residual nuclei, 
i.e. the isotopes of Cr and Ni, while a former trial of this potential 
concerned the proton reaction cross sections $\sigma_R$ \cite{rfc96}. 
Since these data are missing for the Cr nuclei, our local analysis involved 
the isotopes of Mn, Fe, Co, Ni, Cu and Zn, for lower energies important in
statistical emission from excited nuclei. The comparison of these data 
and results of either the local OMP predictions when they are available 
in Table 8 of Ref. \cite{ajk03} or otherwise their proton global OMP 
is shown in Fig. 2. A very good agreement exists apart from the 
isotopes of Fe and in particular Ni, with the data overpredicted by 
about or higher than 10\%. In order to obtain the agreement with the 
corresponding $\sigma_R$ data (Fig. 2) we have found necessary to 
replace the constant real potential diffusivity $a_V$=0.663 fm 
\cite{ajk03} by the energy-dependent forms $a_V$=0.563+0.002$E$ up to 
50 MeV for the target nucleus $^{56}$Fe, and $a_V$=0.463+0.01$E$ up 
to 20 MeV for $^{58}$Ni, where the energy $E$ is in MeV and the 
diffusivity in fm. 
A final validation of both the original OMP and the additional 
energy-dependent $a_V$ has been obtained by analysis of the available 
(p,$\gamma$) and (p,n) reaction data up to $E_p\sim$12 MeV on Cr 
(Fig. 3) and Ni isotopes (Fig. 4) while the other statistical model 
parameters are the same as in the rest of the present work. It can be 
seen that these reaction data have been quite well reproduced, with an 
increase of the related accuracy within 10\% provided by the energy 
dependence adopted for the real potential diffusivity at lower energies.

The  optical potential which is used in this work for calculation 
of the $\alpha$-particle transmission coefficients was established  
previously \cite{va94} for emitted $\alpha$-particles, and supported 
recently by semi-microscopic analysis for $A$$\sim$90 nuclei 
\cite{ma06a}. On the other hand, by comparison of the present 
calculations and measured data \cite{exfor} for the target nuclei 
$^{63,65}$Cu we found that the real well diffuseness $a_R$ of the 
above-mentioned global OMP should be changed to 0.67 fm.
This reduction is rather similar to that found necessary for the 
target nuclei $^{59}$Co, and $^{58,60,62}$Ni \cite{vs04}, so that 
it has been taken into account also for $^{55}$Mn. 
Lastly, the calculation of the deuteron transmission coefficients 
has been carried out by using the global OMP of Lohr and Haeberli 
\cite{jml74} and validated throughout analysis of the 
deuteron-emission spectra at 14.8 MeV \cite{smg79}. 

The back-shifted Fermi gas (BSFG) formula has been used for the 
excitation energies below the neutron-binding energy, with the 
parameters $a$ and $\Delta$ (Table \ref{tab:dens}) obtained by a 
fit of the recent experimental low-lying discrete levels \cite{ensdf} 
and $s$-wave nucleon resonance spacings $D_0$ \cite{ripl2}. Actually 
the same approach basis \cite{chj77,va02,arj98,ajk97} and similar 
parameters have been used as previously within this mass range 
\cite{pr01,vs04}, updated by means of the new structure data 
published in the meantime. Concerning the particle-hole state density 
playing for PE description the same role as the nuclear-level density 
for statistical model calculations, a composite formula \cite{ma98} 
was involved within the GDH model. Thus no s.p.l.-density 
free parameter except for the $\alpha$-particle state density 
\cite{eg81} $g_{\alpha}$=$A/10.36$ MeV$^{-1}$ was used for the PE 
account.

The modified energy-dependent Breit-Wigner (EDBW) model 
\cite{dgg79,ma87} was used for the electric dipole $\gamma$-ray 
strength functions $f_{E1}(E_{\gamma})$ of main importance for 
calculation of the $\gamma$-ray  transmission coefficients, also
as previously within this mass range \cite{pr01,vs04}. The 
corresponding $f_{E1}(E_{\gamma})$ values have been checked within 
the calculations of capture cross sections of Mn and Cu isotopes in 
the neutron energy range from keV to 3-4 MeV, by using the OMP and 
nuclear level density parameters described above and global estimations 
\cite{chj77} of the $\gamma$-ray strength functions for multipoles 
$\lambda\le$3. Thus we found that the $f_{E1}(E_{\gamma})$ strength 
functions corresponding to the experimental \cite{ripl2} average 
radiative widths $\Gamma_{\gamma0}^{exp}$ provide an accurate 
description of the capture data for the Cu isotopes (Fig. 5) while an 
increased value $\Gamma_{\gamma0}\sim$1300 meV has been required in 
the same respect for the $^{55}$Mn nucleus. Finally, the accuracy of 
the $\gamma$-ray strength functions adopted in this work is shown 
also by the above-mentioned analysis of the $(p,\gamma)$ reaction 
cross sections (Figs. 3-4).

\section{Activation cross sections}

The results of the measurements are summarized in Table 1 and are 
shown in Figs. 6-8, using the notation of the preliminary data
in Refs. \cite{aaf99,aaf03}, along with the model calculations. 
The comparison with the previous experimental data concerns in general 
measurements done after the end of '60s. A detailed discussion is 
firstly given below for the new experimental cross sections in 
comparison with previous data as well as the actual calculated 
values.

{\it The $^{55}$Mn(n,2n)$^{54}$Mn reaction cross 
section.} 
Eight new experimental values in the present work cover the energy 
range from 13.56 to 14.78 MeV (Fig. 6). The new data agree within the 
error limits with the earlier measurements, being however higher than 
the most recent previous data from the middle of the last decade but 
systematically lower with respect to the ones measured in the end of 
80's. They seem thus to provide a better confined range of these 
reaction cross sections around 14 MeV, within an accuracy of 3-4\% 
which is most important for the validation of model calculations at 
a similar level. In spite of the smaller weight of PE processes at 
these energies, of around 15\% of the total reaction cross section,
this increased level of accuracy is already acting as a rigorous 
assessment of the model parameters. The rather similar agreement with
these data is a good point for all three calculations, while larger
differences come out just above this energy range. The good agreement 
of the local calculation is mostly due to the local set up of level 
density parameters by fitting of the recent resonance data and 
low-lying level schemes.

{\it The $^{55}$Mn(n,$\alpha$)$^{52}$V reaction cross 
section.}
The nine cross section values measured between 13.47 and 14.83 MeV, 
together with the more recent data of Fessler et al. \cite{af00}, settle 
with enlarged accuracy the maximum of this reaction excitation function. 
It appears to be lower by $\sim$20\% with regard to the data prior 
90's, making thus possible an enhanced knowledge of this excitation 
function. Therefore it results now an obvious change of the slope of 
data up to the incident energy of 12 MeV \cite{mb94}, which is 
described only by one global calculation. The difference of the same 
model prediction with all data above this energy is pointing out the 
less usual trend of the measurement results at the lower energies. 
Since, on the other hand, the agreement found for the 
$^{55}$Mn(n,p)$^{55}$Cr reaction (Fig. 6) between model calculations 
and the measured cross sections also by Bostan and Qaim \cite{mb94} 
is supporting the assumption of a different reaction process leading
to the extra $\alpha$-particle yield at the corresponding excitation 
energies in compound nuclei around 12--18 MeV. The enhancement 
related to the position of a giant quadrupole resonance (GQR) at 
these energies has also been found and discussed for the (n,$\alpha$) 
reaction on $^{92,98}$Mo \cite{ma06a}.

{\it The $^{63}$Cu(n,$\alpha$)$^{60}$Co reaction cross 
section.}
There is a similar case for this reaction as the above-mentioned one
for the target nucleus $^{55}$Mn. The nine cross sections measured 
also between 13.47 and 14.83 MeV remove the ambiguity of about 20\% 
between various sets of measured cross sections at these energies. 
Thus, together with the recent measurement of Plompen et al. 
\cite{exfor,vs04b}, they outline the maximum of this reaction excitation 
function while the latter data set completes this excitation function 
at higher energies (Fig. 7). A previous recent analysis of this 
reaction \cite{vs04b} by using even earlier versions of TALYS and 
EMPIRE-II codes has actually resolved the questions existing on its 
description over the whole energy range. The present global predictions 
of the two codes are rather similar to the former values \cite{vs04b}, 
as well as to the instance of the above-discussed reaction 
$^{55}$Mn(n,$\alpha$)$^{52}$V while the agreement of our local 
calculations is significant especially with reference of the 
simultaneous description of the data for all reaction channels and 
both stable isotopes of copper.

{\it The $^{65}$Cu(n,2n)$^{64}$Cu reaction cross section.}
The data obtained for this reaction are in good agreement with the most
recent experimental data. Moreover, they confirm a sudden change of the 
ascending slope of the excitation function (Fig. 8), which correspond 
in the GDH model to the opening of the partial wave $l$=5 contribution 
for the PE mechanism (see the discussion on Fig. 4 of Ref. \cite{pr05} 
and below).

{\it The $^{65}$Cu(n,p)$^{65}$Ni reaction cross 
section.}
The eight new cross section values of the present work from 13.56 to 
14.78 MeV reduce the spread of the previous measured data from over 
$\sim$40\% (Fig. 8). They are close but slightly lower than the most 
recent measurements \cite{exfor} setting up a strong checking point 
for model calculations. The results of the TALYS code and the 
local approach agree well with them while the EMPIRE-II predictions 
are too large. One may consider at this point that all EMPIRE-II 
calculations in this work made use of the Hybrid Monte-Carlo simulation 
approach for the PE component description, which is the choice 
recommended by the authors \cite{mh04} for neutron energies higher 
than 30 MeV. It can be seen that the actual TALYS-0.72 and EMPIRE-II 
predictions are becoming similar above 20--30 MeV (Figs. 6-8).

The present analysis of the fast-neutron reactions on the stable
isotopes of Mn and Cu involved calculations for the production of 
the ground and isomeric states
$^{52g}$Mn (6$^+$, 5.6 d), 
$^{60m}$Co (2$^+$, 10.5 min), 
$^{62g}$Co (2$^+$, 1.5 min), 
$^{62m}$Co (5$^+$, 13.91 min). 
The production of both low- and high-spin  isomers supports the  
assumptions  adopted,  e.g. within  local-parameter  calculations,  
for the level density angular-momentum distribution as well as the 
$\gamma$-ray strength functions. In the case of $\alpha$-particle 
emission channels this work has additionally validated the 
angular-momentum conservation within the PE model.

\section{Outlook of the pre-equilibrium reactions account}

The formerly-mentioned needs of sound, complete and reliable 
neutron-induced cross section data for Mn and Cu also enabled a 
stringent test of the various nuclear models as well as their
corresponding account of particular effects. However, the key 
points in this respect are related to the PE description, which 
becomes increasingly significant at higher energies. Thus, it seems 
of relevance to look for the answers which may be provided by the 
data analysis firstly below and then above the incident energy of 
20 MeV. Since this is the upper limit of the energy range where 
neutron data are generally considered to be reasonably well known,
the present discussion may also reveal the eventual need for more 
measurements at higher energies. It has been carried out within the 
local approach, based on the use of a consistent parameter set 
already established on the basis of ancillary independent data. 
On the other hand, the insight of the calculated results,
corresponding to distinct parameter values or model assumptions, 
may contribute to the understanding of the variance shown by the
three model calculations in Figs. 6--8. Actually, an ultimate 
goal of this investigation is to increase the global predictions
accuracy to the level of local analysis.

\subsection{Calculated cross-section sensitivity to model parameters, 
below 20 MeV}

\subsubsection{Sensitivity to optical potential parameters}

A first point, following the optical potential analysis described 
in Section 3.2, should concern the effects of the neutron and 
proton modified OMPs on the calculated reaction cross sections.
The larger amount of data existing for the (n,2n) and (n,p) 
reactions on $^{65}$Cu have been involved in this respect as it is 
shown in Fig. 9. Thus, firstly one may note that the modified neutron 
potential (Table 1) is leading to a decrease of $\sim$5\% for the 
(n,2n) reaction calculated cross sections, with respect to the 
results obtained by using the original OMP \cite{ajk03}. 
Similarly, the modified proton potential of Koning and Delaroche leads 
to a decrease of $\sim$20\% of the calculated (n,p) reaction cross 
sections. On the other hand, joining together the two changes in the 
case of the (n,p) reaction, results in a compensation of the latter 
one and a reduced final change of $\sim$10\% for the calculated 
cross sections. Therefore, the additional analysis of the nucleon OMP 
improved the accuracy of the calculated cross sections from $\sim$20\%, 
for the smaller cross sections, to around 5\% for the major reaction 
channels. In fact, this better precision is closer to the 
above-mentioned  range of 3-4\% accuracy of the new measured cross 
sections around 14 MeV, together making possible an effective trial 
of the PE model parameters which are responsible for $\sim$15\% of the 
total reaction cross section at these energies. 


It was noted in the previous section that, by comparison of the 
present calculations and measured data \cite{exfor} for the target 
nuclei $^{63,65}$Cu, it was found that the real well diffuseness $a_R$ 
of the global OMP \cite{va94} for emitted $\alpha$-particles should 
be decreased to 0.67 fm. Since this reduction is rather similar to 
that found recently to be needed for the target nuclei $^{59}$Co and 
$^{58,60,62}$Ni \cite{vs04}, it has been taken into account also for 
the target nucleus $^{55}$Mn. However there are a couple of key 
points related to this matter.
First, by using just the global OMP \cite{va94} for emitted 
$\alpha$-particles, one would obtain also within the local approach 
calculated cross sections for the reaction $^{55}$Mn(n,$\alpha)^{52}$V
closed to the results provided by the code EMPIRE-II and in very good
agreement with the measured cross sections also by Bostan and Qaim 
\cite{mb94}. On the other hand, there is no way to explain the rest
of corresponding data above the incident energy of 12 MeV, by 
pure statistical including PE emission.

Second, Yamamuro \cite{ny96} pointed out, with respect to the clear 
difference of the $\alpha$-particle OMPs which are needed for 
calculation of the ($\alpha$,n) and (n,$\alpha$) reaction cross
sections, that it is found for closed shell nuclei but not for odd 
target nuclei such as $^{53}$Cr, $^{57}$Fe, $^{61}$Ni, and $^{67}$Zn. 
However, the present case of the $^{63,65}$Cu nuclei comes in addition 
of those mentioned in Ref. \cite{vs04}, at variance with Yamamuro's 
statement. Alternatively one may consider the possible enhancement 
related to the position of a giant quadrupole resonance (GQR) at the 
excitation energies concerned in these nuclei. Although generally the 
decay of the GQR is observed with nucleon emission, recent work shows 
\cite{mf05} that an appreciable (non-statistical) decay through 
$\alpha$-particle emission can occur. An extra yield which could
be understood as decay from giant resonances populated via neutron
capture has been found as well for the Mo isotopes \cite{ma06a}.
Thus it follows that further analysis is required, making use also 
of microscopic DF potentials based on temperature-dependent nuclear 
density distributions for the description of (n,$\alpha$) excitation 
functions \cite{ma06a}.

\subsubsection{Charged-particle emission spectra sensitivity}

Actually one may note the same level of 5--20\% differences between the 
global predictions and the measured cross sections, around the incident 
energy of 14 MeV, as the OMP has an effect on the calculated cross 
sections at these energies. Larger divergence occurs at higher energies, 
where it could be related to the continuously growing importance of PE 
assumptions and key quantities.
However, just the assessment within this energy range of the consistent 
parameter set involved in the local approach allows for a further focus 
on differences between the measured and calculated cross sections  
especially above 20 MeV, in order to establish the correctness of the 
adopted PE formalism.

Moreover, in spite of the well-known reduced usefulness of 14 MeV 
neutron reaction data to validate PE calculations for medium-mass nuclei 
\cite{mb83}, the suitable description of related charged-particle 
emission spectra may have a twofold outcome. The lowest-energy region 
of spectra, corresponding to a second emitted particle from a fully 
equilibrated compound nucleus, may truly validate the OMP used for 
emitted particles as well as the level density parameters of the excited 
nuclei. On the other hand, the emission of high-energy charged particles 
is entirely due to the PE processes. Thus, advanced pairing and shell 
corrections of particle-hole state densities could be eventually 
confirmed by the PE model account of this emission-spectrum energy 
region.

Comparison of measured angle-integrated proton and $\alpha$-particle 
emission spectra from 9 \cite{exfor,ct98}, 14.1 \cite{rf85,rf86} and 
14.8 MeV \cite{smg79} neutron-induced reactions on $^{55}$Mn and 
$^{63,65}$Cu  nuclei and calculated values within the local approach 
is shown in Fig. 10. The two goals mentioned above can be considered 
as being satisfied, with a couple of additional comments. While the 
measured particle spectra of Ref. \cite{smg79} are given in the 
laboratory system, their conversion to the center-of-mass system is 
equivalent to a shift of the spectrum to higher energies, of up to 
one MeV for the most energetic $\alpha$-particles (see, e.g., Refs. 
\cite{cd81,rf84}). Thus a good agreement is seen between the measured 
and calculated $\alpha$-particle emission spectra for $^{63,65}$Cu 
nuclei, apart from considerably higher measured data for the high 
energy parts of the spectra corresponding to excitations below 
$\sim$2 MeV in the residual nuclei $^{60,62}$Co. The same effect is 
seen in the case of the target nucleus $^{55}$Mn, for which the 
experimental $\alpha$-particle spectra \cite{rf85,rf86} are given as 
function of channel energy and no further conversion is necessary for 
comparison with the model calculation. This underestimation was noted 
as well for other target nuclei in this mass region \cite{cd81,rf84},  
indicating that there may be considerable direct excitation of 
residual nuclei low lying levels beyond the validity of the PE models.

Concerning the additional underestimation of the $\alpha$-particle 
spectrum on $^{55}$Mn at lowest energies, one should note that the 
measurements are complicated in this energy range by a rather 
large background \cite{rf84}. 

\subsection{Model-analysis support above 20 MeV}

Above the neutron incident energy of 20 MeV, particularly at 
excitation energies beyond 30 MeV, the PE model becomes increasingly 
important in determining the reaction cross section. The lack of 
free parameters within the corresponding GDH model, as well as the 
consistent use of the same optical potential and nuclear level 
density parameters as the HF model, make possible a focus on the 
correctness of the main related quantity which is the particle-hole 
state density. 

\subsubsection{Nuclear potential finite-depth correction}

The original GDH formalism \cite{mb83,mb73} considered a Fermi 
distribution for the nuclear matter density, with the Fermi energy 
$F$=40 MeV at saturation density. On this basis it takes into 
account the nuclear surface effects by means, firstly, of the sum of 
contributions due to different entrance channel partial waves $l$ 
for the first projectile-target interaction. The relevant parameters 
in this case are averaged over the nuclear densities corresponding 
to the entrance-channel trajectories from a point at which the
nuclear density is $\sim$1/150 of its saturation value to the 
radius $R_l$=$^{\prime}\!\!\!\lambda$$(l+\frac{1}{2})$. 
Secondly, lower local-density Fermi energies \cite{eg73} calculated 
for each of these trajectories, $F_1(R_l)$, have been considered 
within the particle-hole state densities (PSD) and limited the hole 
degrees of freedom. They correspond to a finite well depth correction 
which has been included \cite{ck85} in the PSD equidistant spacing 
model at the same time as the advanced pairing \cite{ck87} and shell 
corrections \cite{cyf84} added to the Pauli correction, and the 
non-equidistant single-particle levels \cite{mh89}. All of the above
were included in a PSD composite formula \cite{ma98} added to the 
GDH model within the STAPRE-H code, and were part of previous studies 
carried out in a similar way \cite{pr01,vs04,pr05,va06} but at 
incident energies up to 20 MeV. The extension of the present analysis 
to 40 MeV, by means of the measured data put together in Figs. 11-12, 
is able to check the importance of the finite-depth correction in the 
frame of the PSD composite formula \cite{ma98}. 

Thus, the vanishing of this correction is obtained by replacing the
local-density Fermi energies $F_1(R_l)$ with the Fermi energy central 
value $F$=40 MeV. The results of this exercise are shown in Fig. 11, 
the most apparent and direct view corresponding to the 
$^{65}$Cu(n,p)$^{65}$Ni reaction. The GDH $l$-dependent finite-depth 
corrections $F_1(R_l)$ allow the opening only with the energy increase 
of the PE contribution due to each higher partial wave. This attribute, 
together with the decreasing total reaction cross section $\sigma_R$ 
with energy increasing, leads to a rising fraction $\sigma_{PE}$/$\sigma_R$ 
but a rather constant (n,p) reaction cross section above the incident 
energy $\sim$20 MeV where the emission of a second neutron becomes
possible. By raising the local-density Fermi energies to the central
value, the PE contributions of all partial waves become possible from 
the beginning, in the limit set by the corresponding transmission 
coefficients. Thus the fraction $\sigma_{PE}$/$\sigma_R$ will increase
faster while, e.g., the (n,p) reaction cross section will decrease 
continuously after getting a higher maximum. The latter two attributes
are both opposed to the experimental data, even if their energy
dependence above 20 MeV is only fairly accurate. The same findings
follow the analysis of the other data above 20 MeV shown in Fig. 11,
as well as the reaction $^{65}$Cu(n,2n)$^{64}$Cu added for completion.
Changes of the calculated reaction cross sections above the incident
energy of 20 MeV, corresponding to this finite-depth correction, are 
going from $\sim$50\% to more than 100\%. 
A similar result was noticed by Korovin et al. \cite{yak01}, within a 
modified GDH model and using the former PSD formula of Ericson, while 
the present results are based on the composite formula \cite{ma98}.
However, excepting the (n,4n) and (n,p$\alpha$) reactions on $^{55}$Mn 
and the (n,3n) reaction on $^{63}$Cu, the need for more accurate 
measured data at least up to 40 MeV is obvious.

\subsubsection{Single-particle levels density effects}

The Fermi-gas model (FGM) energy dependence of the s.p.l. density 
has been used within the PSD composite formula \cite{ma98}, in the 
present local approach as well as in the recent similar analyses 
\cite{pr01,vs04,pr05,va06}, following the study and conclusions of 
Herman et al. \cite{mh89}. Actually average values of the s.p.l. of
excited particles and holes, $g_p(p,h)$=$g(F+{\overline u_p})$ and
$g_p(p,h)$=$g(F-{\overline u_h})$ respectively, have been obtained 
corresponding to the average excitation energies for particles and 
holes ${\overline u_p}$ and and ${\overline u_h}$ \cite{ma98}. The 
s.p.l. density value at the Fermi level has been derived on the basis 
of its relation to the nuclear level density parameter,  
$g(F)$=$\frac{6}{\pi^2}a$, by using the parameter values given in 
Table \ref{tab:dens}. By replacing the above-mentioned average 
energy-dependent s.p.l. $g_{p(h)}(p,h)$ with the constant value 
$g(F)$ result in the changes shown in Fig. 11. They are quite small 
below 20 MeV, increasing for the incident energies up to 40 MeV from 
$\sim$5\% to 20\%.

Two features should be pointed out at this time. Firstly, the energy 
dependence of the s.p.l. density is much less important than its value 
at the Fermi energy \cite{va06}. This is a consequence of the fact 
that the PE cross section is determined by a ratio of the particle-hole 
level densities corresponding to exciton  configurations which differ 
by one excited particle \cite{mb83,mb73,ma94}. Secondly, this change 
may become significant at energies higher than 40 MeV.

\subsubsection{Nuclear-shell effects}

The PSD composite formula \cite{ma98} included the advanced pairing 
\cite{ck87} and shell corrections \cite{cyf84} added to the Pauli 
correction, by taking into account the nuclear-shell effects through
an additional back-shift $S$ of the effective excitation energy 
\cite{cyf84}. It has been connected, for the excitation energies 
lower than the binding energy, to the BSFG virtual ground-state 
shift parameter $\Delta$=$U_p$+$S$, where the former term is a 
constant pairing correction corresponding to the PSD closed formula 
\cite{cyf84}. The washing out of shell effects above the neutron binding 
was taken into account also for the back-shift $S$ value by using the 
shell correction within the approach of Junghans et al. \cite{arj98}, 
derived as mentioned in Section 3.2.3, with a similar smooth transition 
between the two energy range as for the nuclear level density. Obviously,
the largest effect of this PSD correction corresponds to the lowest
excitation energies, the related $S$-values causing up the high-energy 
limit of the emitted-particle spectra (e.g., Fig. 10). Since the shift 
$\Delta$ is around and less than zero value for the odd-$A$ and odd-odd 
nuclei, respectively, the back-shift $S$ is negative for most of the 
nuclei involved in the present work. It leads to enlarged effective 
excitation energies for the PSD calculation at lower excitations, 
finally increasing the PE cross sections. In Fig. 11 is also shown the 
effect of removing the shell correction in the PSD composite formula, 
the subsequent decrease of the PE component leading to, e.g., (n,2n) 
and (n,3n) reaction cross sections decreased by 5--10\% around the 
maximum of their excitation functions but (n,p) reaction cross 
sections which decrease by $\sim$20\%, around the incident energy of 
14 MeV, up to more than 50\% at 40 MeV. The effect is obviously less 
important for (n,4n) reactions, at least at the energies involved in 
this work, where multiple PE processes are not important \cite{ajk04}. 
Nevertheless, consideration of the nuclear-shell effects proves to be
quite important within the present analysis, in addition to the 
influence of the $f_{7/2}$ neutron and proton shell closures on the 
particle PE spectra \cite{mb85} already noted.

\subsubsection{The s.p.l.--density 'continuum effect'}

The {\it 'continuum effect'} (CE), i.e. the s.p.l. density decreasing 
with energy in the continuum region \cite{ss92,yab96,ss97}, can be 
described basically using the corrected s.p.l. density formula 
\cite{ma98}

\begin{equation}\label{eq:25}
 g_p(p,h) = g(F) 
  \left[\left(1+{{\overline u_p}\over{F}}\right)^{1/2}-
  \left({{{\overline u_p}-B}\over{F}}\right)^{1/2} 
  \theta({\overline u_p}-B)\right] \: ,
\end{equation}
where $B$ is the nucleon binding energy. However, for the present 
reaction cross section calculation, one should also take into  
account the Coulomb and centrifugal barriers (e.g., \cite{ob81}). 
On the other hand, one should note that the role of this effect will 
be major in the case of the particle-hole configurations of the 
composite nucleus, excited at higher energies with respect to the 
residual nuclei. 

The progressive addition of these barriers to the binding energy $B$, 
as well as the removal of the continuum effect within the PSD 
calculation, are shown in Fig. 12. The most apparent and direct view 
can be seen once more in the reaction $^{65}$Cu(n,p)$^{65}$Ni. The 
decrease of the s.p.l. densities  due to the consideration of binding 
energy alone, with respect to no CE presumed, leads to the increase 
of PE cross sections. The addition of the Coulomb barrier actually 
decreases the CE weight, which remains visible only above the incident 
energy of $\sim$25 MeV. Finally, the inclusion also of the centrifugal 
barrier reduces even more the CE size within the energy range discussed 
in the present work. The CE complete treatment may again play an 
important role at higher energies, the steps of its partial account in 
this analysis being able to shed some light on the expected 
consequences at these energies.

\section{Summary and conclusions}

New measurements with the activation technique were performed for 
neutron-induced reactions around 14 MeV on the stable isotopes of 
Mn and Cu. A significant body of experimental data with an 
accuracy within 3--4\% has been obtained by measurements on 
natural samples for the    
$^{55}$Mn(n,2n)$^{54}$Mn, 
$^{55}$Mn(n,$\alpha$)$^{52}$V, 
$^{63}$Cu(n,$\alpha$)$^{60}$Co, 
$^{65}$Cu(n,2n)$^{64}$Cu, and
$^{65}$Cu(n,p)$^{65}$Ni reactions  from 13.47 to 14.83 MeV. The 
experimental cross sections are compared with the results of 
calculations for all activation channels for $^{55}$Mn and 
$^{63,65}$Cu isotopes, and neutron incident energies below 20 
MeV as well as up to 40--50 MeV. The increased accuracy of 
the present cross sections around 14 MeV has made possible an 
effective trial of the PE model parameters  at the same level, 
even if this reaction mechanism is responsible at these energies 
for only $\sim$15\% of the total reaction cross section. 
It should be also noted that similar differences of 5--20\% 
exist between the global predictions and the measured cross 
sections, in the same energy range, as the OMP effects on the 
calculated cross sections. On the other hand, this assessment 
of the consistent parameter set involved in the local analysis 
below 20 MeV allows a further focus at higher energies on 
differences between the measured and calculated cross sections 
related to model assumptions. 
The few more recent data of increased accuracy between $\sim$14 
and 20--21 MeV are also quite useful in this respect.

Larger divergence between the measured and calculated cross
sections occurs mainly for the global predictions at higher 
energies, where the importance of PE assumptions and key 
quantities is continuously increasing. Since an ultimate goal
of this investigation is to increase the global prediction
accuracy to the level of local analysis, we have looked for the
significant effects related to distinct parameter or model 
assumptions. 
The most important is found to correspond within the GDH model
to the nuclear potential finite-depth correction taken into 
account for description of particle-hole state densities. Its 
omission leads to a large increase of the PE weight as well as
to reaction cross section changes going from $\sim$50\% to more 
than 100\%. However, the need for more accurate measured data at 
least up to the incident energy of 40 MeV is obvious. A similar 
case is shown by consideration of the nuclear-shell effects
within the PSD formula. On the other hand, there are effects 
such as the s.p.l.--density energy dependence and inclusion of the 
{\it 'continuum effect'} which may however become significant at 
energies higher than 40 MeV. Therefore, the present discussion 
is also pointing out the usefulness of further measurements of
neutron activation reactions at higher incident energies below,
e.g., 40 MeV \cite{xl06} as well as up to 100 MeV \cite{am05}.

\section*{Acknowledgments}
Work supported in part by the European Community EFDA under the 
Contract of  Association EURATOM--MEdC (Bucharest), and MEdC 
Contract No. CEEX-05-D10-48 .

\newpage
\section*{Figure captions}

\noindent
FIG. 1. Comparison of experimental and calculated neutron total
cross  sections for $^{55}$Mn and $^{63,65,nat}$Cu target nuclei, 
by using the global (dotted curves) and local (dashed curves) OMP 
parameter sets of Koning and Delaroche \cite{ajk03}, and the changes
of the latter given in Table 1 (solid curves). The experimental data 
are taken from the EXFOR database \cite{exfor}.

\noindent
FIG. 2. Comparison of the measured \cite{rfc96} and calculated  
proton reaction cross sections on all stable isotopes of Mn, 
Fe, Co, Ni, Cu and Zn, by using either the local OMP 
predictions of Koning and Delaroche when they are available in 
Table 8 of Ref. \cite{ajk03} or otherwise their proton global OMP 
(dotted, dash-dotted and dashed curves) and the modified parameter
set mentioned in the text for the target nuclei $^{56}$Fe and 
$^{58}$Ni (solid curves).

\noindent
FIG. 3. Comparison of the measured \cite{exfor} and calculated  
proton reaction cross sections (dash-dotted curve), (p,$\gamma$) 
and (p,n) reaction cross sections up to $E_p\sim$12 MeV on Cr 
isotopes by using the OMP parameter sets mentioned for Fig. 2.

\noindent
FIG. 4. As for Fig. 3, but for the Ni isotopes.

\noindent
FIG. 5. Comparison of the measured \cite{exfor} neutron-capture  
cross sections of $^{55}$Mn and $^{63,65,nat}$Cu target nuclei,
for incident energies up to 3--4 MeV, and calculated values by using 
the computer codes TALYS-0.72 (dashed curves) and EMPIRE-II 
(dash-dotted curves) with default global parameters, and the local
analysis with $\gamma$-ray strength functions $f_{E1}(E_{\gamma}$) 
within the EDBW model corresponding to either the experimental 
\cite{ripl2} average $s$-wave radiative widths $\Gamma_{\gamma0}$ 
(dotted curves), or $\Gamma_{\gamma0}$ values corresponding to a fit 
of experimental neutron capture data (solid curves).

\noindent
FIG. 6. Comparison of measured \cite{exfor} and calculated  
neutron-activation cross sections for the target nucleus $^{55}$Mn, 
by using the computer codes TALYS-0.72 (dashed curves) and EMPIRE-II 
(dash-dotted curves) with default global parameters, and STAPRE-H 
(solid curves) with the local parameter set given in this work.

\noindent
FIG. 7. As for Fig. 6, but for the target nucleus $^{63}$Cu.

\noindent
FIG. 8. As for Fig. 6, but for the target nucleus $^{65}$Cu.

\noindent
FIG. 9. Comparison of measured \cite{exfor} and calculated cross sections 
within the local approach for the (n,2n) and (n,p) reactions on the 
target nucleus $^{65}$Cu, by using the OMP parameter sets of Koning and 
Delaroche \cite{ajk03} (dotted curves), and corresponding changes for the proton OMP (dashed curve) and neutron OMP (solid curves).

\noindent
FIG. 10. Comparison of measured \cite{exfor,smg79,ct98,rf85,rf86} 
angle-integrated proton and $\alpha$-particle emission spectra from 9, 14.1 
and 14.8 MeV neutron-induced reactions on the $^{55}$Mn and $^{63,65}$Cu 
nuclei and calculated values within the local approach for the PE emission 
(dashed curves), statistical first- (dash-dotted curves) and second-emitted 
particles (dotted curves) from equilibrated compound nuclei, and their sum 
(solid curves).

\noindent
FIG. 11. Comparison of measured \cite{exfor} neutron-activation cross 
sections for the target nuclei $^{55}$Mn and $^{63,65}$Cu up to 40 MeV, 
and calculated values with the local parameter set given in this work
(solid curves) except for replacement of either the local-density Fermi 
energies $F_1(R_l)$ with the Fermi energy central value $F$ (dash-dotted
curves), or the average energy-dependent s.p.l. densities with the 
constant value $g(F)$ (dotted curves), as well as for removal of the 
shell correction $S$ in the PSD composite formula (dashed curves).

\noindent
FIG. 12. As for Fig. 11, but for removal of the {\it 'continuum effect'} 
(CE) of the s.p.l. density within the particle-hole state density 
calculation (dotted curves), and taking into account for this effect the 
nucleon binding energy $B$ either alone (dash-dotted curves) or together 
with the Coulomb barrier $B_C$ (dashed curves), while the solid curves 
correspond to consideration of also the centrifugal barrier $B_{CF}$.

\newpage

\begin{table*} 
\caption{\label{tab:xsec}Measured reaction cross sections (mb) for  
$^{55}$Mn and $^{63,65}$Cu isotopes between 13.5 and 14.8 MeV. Mean 
and full widths at half maximum (fwhm) of the neutron energy 
distribution are shown. The uncertainty of the mean energy is 
10 keV. Standard uncertainties are given for the cross sections.}
\bigskip

\begin{tabular}{cccccc}
\hline
Energy& \multicolumn{5}{c}{Reaction}\\
\cline{2-6}
 (MeV) & $^{55}$Mn(n,2n)$^{54}$Mn & $^{55}$Mn(n,$\alpha$)$^{52}$V &
   $^{63}$Cu(n,$\alpha$)$^{60}$Co & $^{65}$Cu(n,2n)$^{64}$Cu      &
   $^{65}$Cu(n,p)$^{65}$Ni\\ 
\hline

13.47 &         & 20.6(8) & 47.5(27)&         &          \\
13.56 & 620(26) &         &         & 834(49) & 22.33(76)\\
13.65 &         & 20.9(9) & 45.6(16)&         &          \\
13.74 & 632(26) &         &         & 863(37) &20.28(127)\\
13.88 &         & 22.6(10)& 45.6(25)&         &          \\
13.96 & 656(27) &         &         & 918(62) & 21.45(91)\\
14.05 &         & 22.7(10)& 45.5(14)&         &          \\
14.10 & 690(25) & 22.2(10)& 45.3(8) & 888(25) & 21.3(7)  \\
14.19 & 708(27) &         &         & 867(37) &20.45(116)\\
14.27 &         & 22.9(10)& 45.8(14)&         &          \\
14.42 & 740(28) &         &         & 952(43) & 20.36(71)\\
14.44 &         & 23.3(9) & 46.1(17)&         &          \\
14.61 & 763(27) &         &         & 903(38) & 21.46(65)\\
14.63 &         & 23.5(9) & 43.2(15)&         &          \\
14.78 & 781(28) &         &         & 965(50) & 20.58(87)\\
14.83 &         & 22.8(10)& 42.3(13)&         &          \\

\hline
\end{tabular}
\end{table*}

\newpage
\begin{table*} 
\caption{\label{tab:nres}Comparison of experimental \cite{ripl2} and 
calculated neutron scattering parameters of $^{55}$Mn and $^{63,65}$Cu 
isotopes at the neutron energies of 100, 80 and 50 keV, respectively,
and (bottom) the changes of OMP parameters \cite{ajk03} which provide 
the best SPRT results, where the energies are in MeV and geometry 
parameters in fm.}

\begin{tabular}{l|ccc|ccc|ccc} 
\hline
\hspace*{0.5in} $\backslash \:$ Target & \multicolumn{3}{c|}{$^{55}$Mn}
       & \multicolumn{3}{|c|}{$^{63}$Cu}
       & \multicolumn{3}{|c}{$^{65}$Cu} \\
\cline{2-4} \cline{5-7} \cline{8-10} 
Potential & S$_0$*10$^4$ & S$_1$*10$^4$ & R' & 
            S$_0$*10$^4$ & S$_1$*10$^4$ & R' & 
            S$_0$*10$^4$ & S$_1$*10$^4$ & R' \\
\hline
Exp. &
 4.4(6) & 0.3(1) & & 2.1(3) & 0.44(7) &  & 2.2(3) & 0.47(8) & \\
Ref. \cite{ajk03} - global &
  3.8  & 0.70 & 6.2 & 2.2 &  0.81  &  7.0  &  2.1   &  0.81  &  7.4 \\
Ref. \cite{ajk03} - local &
  4.1  & 0.58 & 6.2 & 2.1 &  0.77  &  7.1  &  1.76  &  0.75  &  7.4 \\
Ref. \cite{ajk03} - local +&
  3.8  & 0.48 & 4.6 & 2.2 &  0.48  &  6.0 &  1.92  &  0.48  &  6.8 \\
\cline{2-4} \cline{5-7} \cline{8-10} 
\hspace*{0.55in} changes: 
       & \multicolumn{3}{c|}{ }
       & \multicolumn{3}{|c|}{$r_V$$=$1.260-0.02$E$, $E$$<$3}
       & \multicolumn{3}{|c}{$r_V$$=$1.251-0.016$E$, $E$$<$3} \\
       & \multicolumn{3}{c|}{$a_V$$=$0.563+0.02$E$, $E$$<$5}
       & \multicolumn{3}{|c|}{$a_V$$=$0.213+0.15$E$, $E$$<$3}
       & \multicolumn{3}{|c}{$a_V$$=$0.303+0.12$E$, $E$$<$3} \\
\hline
\end{tabular}
\end{table*}

\clearpage                               
\noindent
\begin{longtable}{cccccccccc}                                 
\caption{\label{tab:dens} The low-lying levels number $N_d$ up to excitation 
energy $E_d$ \protect\cite{ensdf} used in Hauser-Feshbach calculations, 
and the low-lying levels and $s$-wave nucleon-resonance spacings 
$D_0^{exp}$ (Ref. \cite{ripl2} except otherwise noted)
in the nucleon energy range $\Delta$E above the respective 
binding energy $B$, for the target-nucleus ground-state spin $I_0$, 
fitted in order to obtain the BSFG level-density parameter {\it a} and 
ground-state shift $\Delta$ (corresponding to a spin cut-off factor 
calculated with a variable moment of inertia between the half and 75\% 
of the rigid-body value, for the excitation energies from g.s. to the 
nucleon binding energy, and the reduced radius r$_0$=1.25 fm).}\\

\hline
Nucleus   &$N_d$&$E_d$& \multicolumn{5}{c}
                     {Fitted level and resonance data}& $a$ &$\Delta$ \\
\cline{4-8}
           &  &      &$N_d$&$E_d$&$B+\frac{\Delta E}{2}$&
                                     $I_0$&$D_0^{exp}$ \\ 
           &  & (MeV)&   & (MeV)& (MeV)&     &(keV)&(MeV$^{-1}$)&(MeV)\\ 
\hline
\endfirsthead
\caption{continued.}\\
\hline
Nucleus   &$N_d$&$E_d$& \multicolumn{5}{c}
                     {Fitted level and resonance data}& $a$ &$\Delta$ \\
\cline{4-8}
           &  &      &$N_d$&$E_d$&$B+\frac{\Delta E}{2}$&
                                     $I_0$&$D_0^{exp}$ \\ 
           &  & (MeV)&   & (MeV)& (MeV)&     &(keV)&(MeV$^{-1}$)&(MeV)\\ 
\hline
\endhead
\hline
\endfoot
$^{50}$Ti&19&4.940&19&4.94&11.059&7/2& 4.0(8)     & 5.55& 1.20 \\
$^{51}$Ti&22&4.187&18&3.77& 6.565& 0 & 125(70)    & 6.05& 0.40 \\
$^{47}$V &23&2.810&23&2.81& 7.750& 0 &36.0(48)$^a$& 5.50&-1.15 \\
$^{48}$V &23&1.781&23&1.78&      &   &            & 5.95&-1.85 \\
$^{49}$V &25&2.408&25&2.41& 9.559& 0 &10.6(10)$^a$& 5.35&-1.80 \\
$^{50}$V &32&2.162&46&2.65& 7.361&7/2& 4.1(6)     & 5.95&-1.75 \\
$^{51}$V &37&3.683&54&4.12&10.646& 0 & 7.9(6)$^a$ & 5.65&-0.68 \\
         &  &     &  &    &11.071& 6 & 2.3(6)                  \\
$^{52}$V &20&1.843&20&1.84& 7.361&7/2& 4.1(6)     & 6.15&-1.60 \\
$^{53}$V &25&2.967&25&2.97&      &   &            & 5.65&-1.03 \\
$^{54}$V &19&1.752&17&1.54&      &   &            & 5.95&-1.85 \\
$^{50}$Cr&32&4.363&32&4.36&      &   &            & 5.40& 0.00 \\
$^{51}$Cr&41&3.448&85&4.29& 9.561& 0 &13.3(13)$^a$& 5.50&-1.20 \\
$^{52}$Cr&17&4.100&17&4.10&      &   &            & 5.55& 0.20 \\
$^{53}$Cr&31&3.617&27&3.44& 8.432& 0 &43.40(437)  & 5.35&-0.90 \\
$^{54}$Cr&33&4.458&33&4.46& 9.817&3/2&  7.8(8)    & 5.55& 0.10 \\
$^{55}$Cr&24&2.895&24&2.90& 6.696& 0 & 54.4(8)$^a$& 6.02&-0.82 \\
         &  &     &  &    &      &   & 62.0(8)                 \\
$^{50}$Mn& 6&1.143& 6&1.14&      &   &            & 5.85&-1.40 \\
$^{51}$Mn&20&2.984&28&3.29&      &   &            & 5.55&-0.85 \\
$^{52}$Mn&20&2.337&17&2.13&      &   &            & 6.00&-1.20 \\
$^{53}$Mn&36&3.555&42&3.73&      &   &            & 5.35&-1.10 \\
$^{54}$Mn&24&1.925&24&1.93&      &   &            & 6.05&-1.81 \\
$^{55}$Mn&32&2.953&45&3.07&10.497& 0 & 7.1(7)$^a$ & 5.70&-1.55 \\
$^{56}$Mn&23&1.384&37&1.88& 7.374&5/2& 2.3(4)     & 6.10&-2.30 \\
$^{57}$Mn&21&2.233&21&2.33&      &   &            & 6.20&-1.25 \\ 
$^{58}$Mn&11&0.882&11&0.88&      &   &            & 6.65&-1.95 \\ 
$^{58}$Fe&42&4.350&60&4.72&10.139&1/2& 6.5(10)    & 6.15& 0.15 \\ 
$^{59}$Fe&28&2.856&28&2.86& 6.755& 0 &25.4(49)    & 6.70&-0.70 \\ 
$^{60}$Fe&21&3.714&21&3.71&      &   &            & 6.15& 0.15 \\ 
$^{61}$Fe& 3&0.391&13$^b$&1.75&  &   &            & 6.85&-1.00 \\ 
%
$^{55}$Co&23&3.775&23&3.78&      &   &            & 5.35&-0.40 \\ 
$^{56}$Co&28&2.969&20&2.61&      &   &            & 6.20&-0.78 \\ 
$^{57}$Co&34&3.296&70&4.11& 8.819& 0 &19.4(24)$^a$& 5.75&-0.98 \\ 
         &  &     &  &    & 9.591&   &13.3(11)$^a$             \\
$^{58}$Co&29&1.606&41&1.93&      &   &            & 6.50&-2.23 \\ 
$^{59}$Co&38&3.090&68&3.67&10.217& 0 & 4.3(4)$^a$ & 6.40&-0.85 \\ 
$^{60}$Co&35&1.833&41&1.98& 7.542&7/2& 1.25(15)   & 6.95&-1.70 \\ 
$^{61}$Co&24&2.499&28&2.64&      &   &            & 6.85&-0.75 \\ 
$^{62}$Co&12&0.920&16&1.27&      &   &            & 7.30&-1.55 \\ 
$^{63}$Co&11&2.191&11&2.19&      &   &            & 7.30&-0.30 \\ 
$^{64}$Co& 8&0.953&17$^b$&1.36&  &   &            & 7.75&-1.30 \\ 
$^{57}$Ni&22&4.374&22&4.37&      &   &            & 5.70& 0.46 \\ 
$^{58}$Ni&32&4.752&32&4.75&      &   &            & 5.90& 0.65 \\ 
$^{59}$Ni&36&3.196&57&3.64& 9.411& 0 &13.4(9)     & 5.90&-1.10 \\ 
         &  &     &  &    &      &   &12.5(9)$^a$              \\  
$^{60}$Ni&31&4.116&45&4.54&11.394&3/2& 2.0(7)$^a$ & 6.10& 0.20 \\ 
$^{61}$Ni&21&2.129&21&2.13& 8.045& 0 &13.8(9)     & 6.40&-1.24 \\ 
         &  &     &  &    &      &   &13.9(15)$^a$&     &      \\
$^{62}$Ni&25&3.860&47&4.46&10.631&3/2&2.10(15)    & 6.40& 0.27 \\ 
$^{63}$Ni&19&2.353&19&2.35& 7.117& 0 &  16(3)     & 7.35&-0.52 \\ 
         &  &     &  &    & 7.238&   &  15(2)$^a$ &     &      \\
$^{64}$Ni&29&4.285&49&4.76&      &   &            & 6.85& 0.79 \\ 
$^{65}$Ni&20&2.520&20&2.52& 6.398& 0 &19.6(30)    & 7.80&-0.20 \\ 
$^{59}$Cu&15&2.715&15&2.72&      &   &            & 6.25&-0.45 \\ 
$^{60}$Cu&17&1.007&22&1.43&      &   &            & 7.00&-1.75 \\ 
$^{61}$Cu&35&3.092&35&3.09&      &   &            & 6.55&-0.67 \\ 
$^{62}$Cu&18&1.077&56&1.92&      &   &            & 7.10&-2.00 \\ 
$^{63}$Cu&36&2.978&40&3.10& 9.026& 0 & 5.9(7)$^a$ & 7.08&-0.50 \\ 
$^{64}$Cu&45&1.918&84&2.42& 7.993&3/2&0.95(9)     & 7.25&-1.78 \\ 
$^{65}$Cu&21&2.669&51&3.36&      &   &            & 7.70&-0.15 \\ 
$^{66}$Cu&22&1.439&22&1.44& 7.166&3/2&1.30(11)    & 7.95&-1.35 \\ 
%
\hline
\end{longtable}
$^a$Ref. \cite{hv88}\\
$^b$Levels of similar isotope in the close neighbouring.\\

\newpage

\newpage
\begin{figure}[t]	
\centerline{\epsfig{file=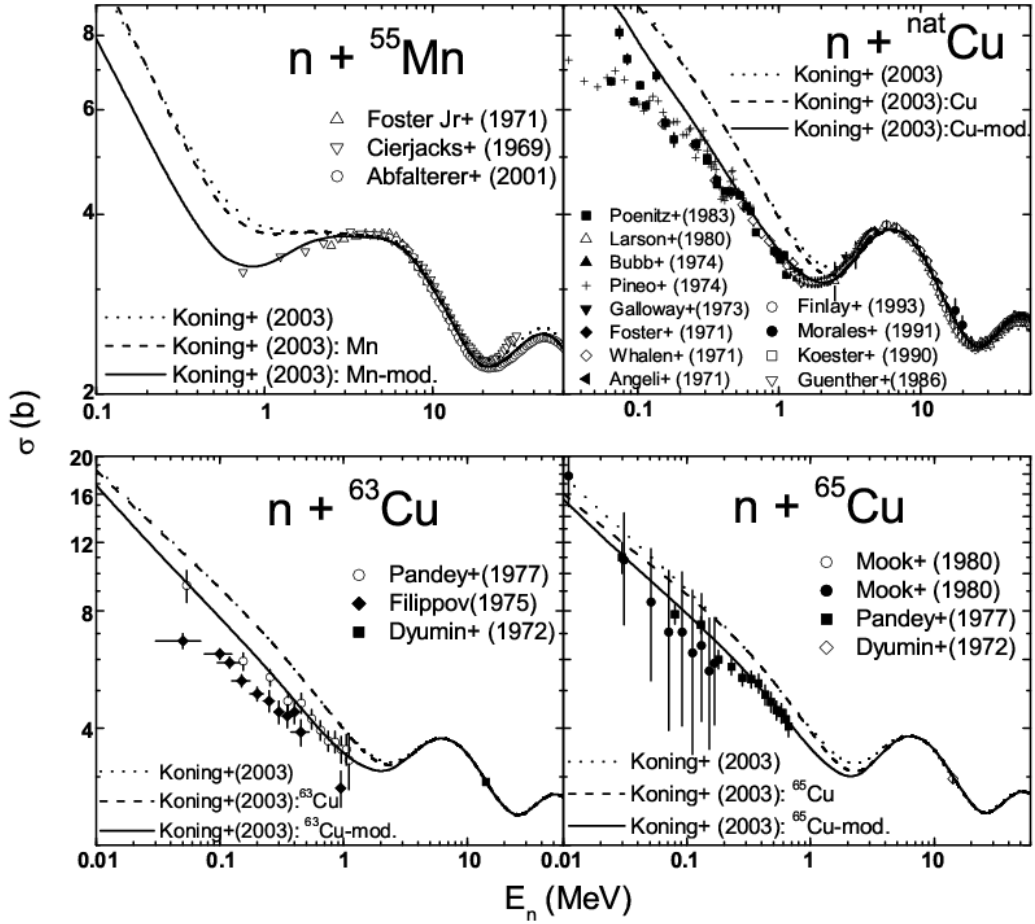,height=14.cm}}
\caption{Comparison of experimental and calculated neutron total
cross  sections for $^{55}$Mn and $^{63,65,nat}$Cu target nuclei, 
by using the global (dotted curves) and local (dashed curves) OMP 
parameter sets of Koning and Delaroche \cite{ajk03}, and the changes
of the latter given in Table 1 (solid curves). The experimental data 
are taken from the EXFOR data basis \cite{exfor}.}
\end{figure}

\newpage
\begin{figure}[t]	
\centerline{\epsfig{file=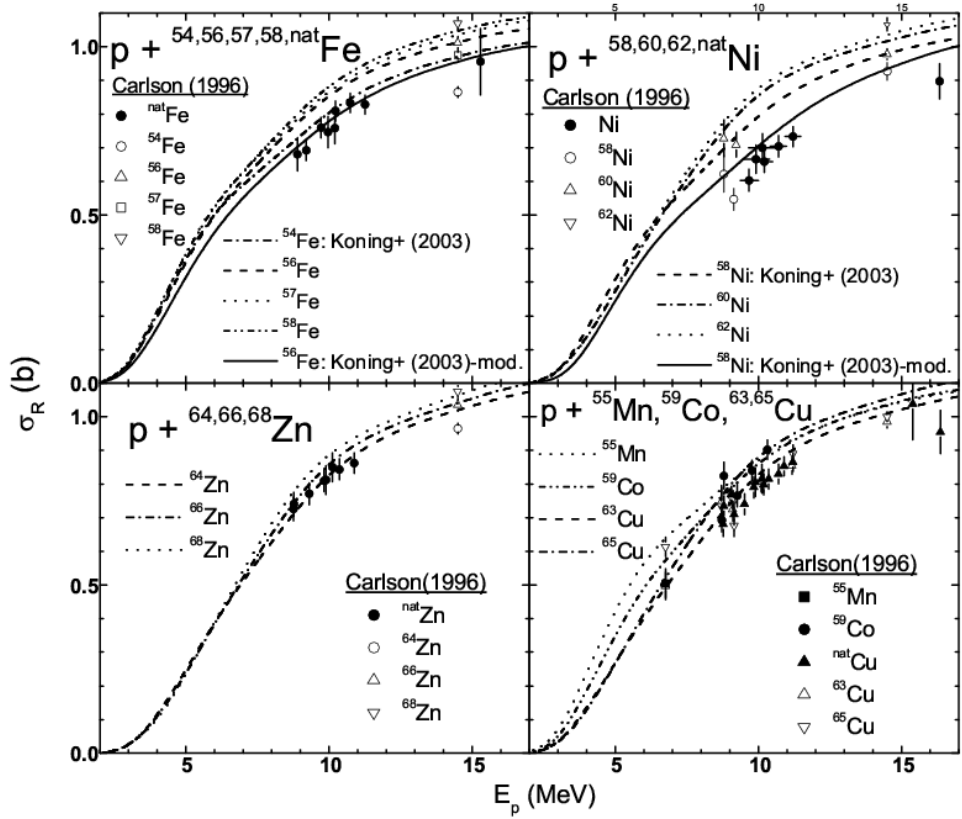,height=12.5cm}}
\caption{Comparison of the measured \cite{rfc96} and calculated  
proton reaction cross sections on all stable isotopes of Mn, 
Fe, Co, Ni, Cu and Zn, by using either the local OMP 
predictions of Koning and Delaroche when they are available in 
Table 8 of Ref. \cite{ajk03} or otherwise their proton global OMP 
(dotted, dash-dotted and dashed curves) and the modified parameter
set mentioned in the text for the target nuclei $^{56}$Fe and 
$^{58}$Ni (solid curves).}
\end{figure}

\newpage
\begin{figure}[t]	
\centerline{\epsfig{file=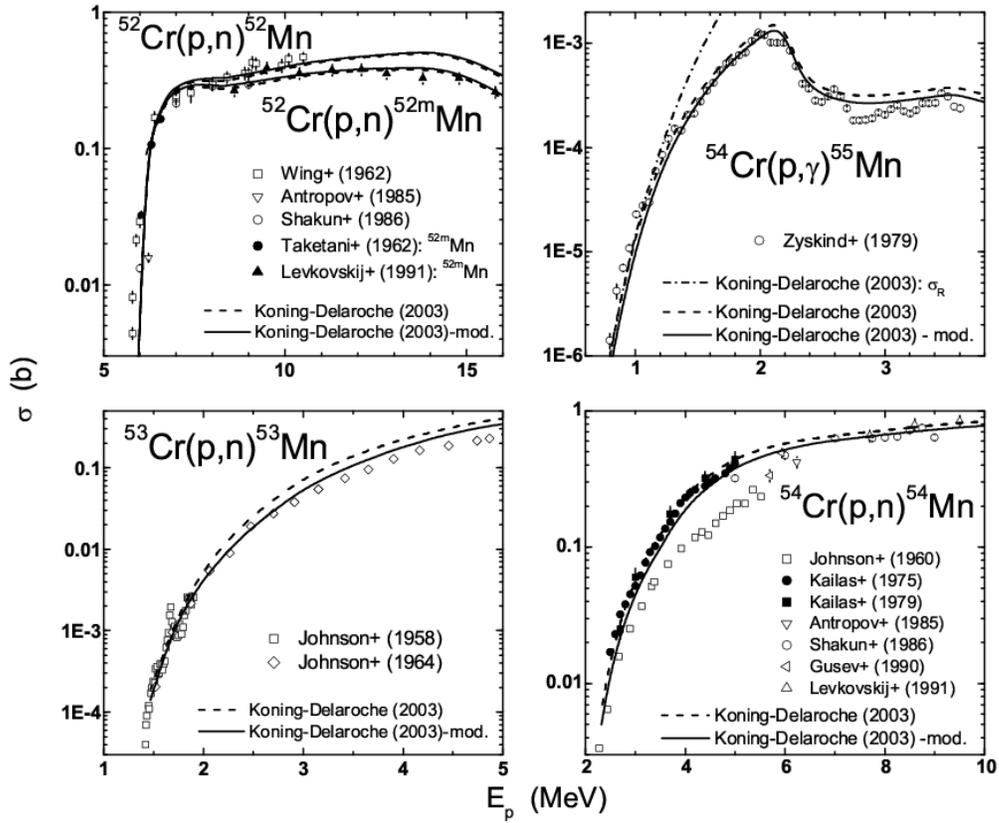,height=12.5cm}}
\caption{Comparison of the measured \cite{exfor} and calculated  
proton reaction cross sections (dash-dotted curve), (p,$\gamma$) 
and (p,n) reaction cross sections up to $E_p\sim$12 MeV on Cr 
isotopes by using the OMP parameter sets mentioned for Fig. 2.}
\end{figure}

\newpage
\begin{figure}[t]	
\centerline{\epsfig{file=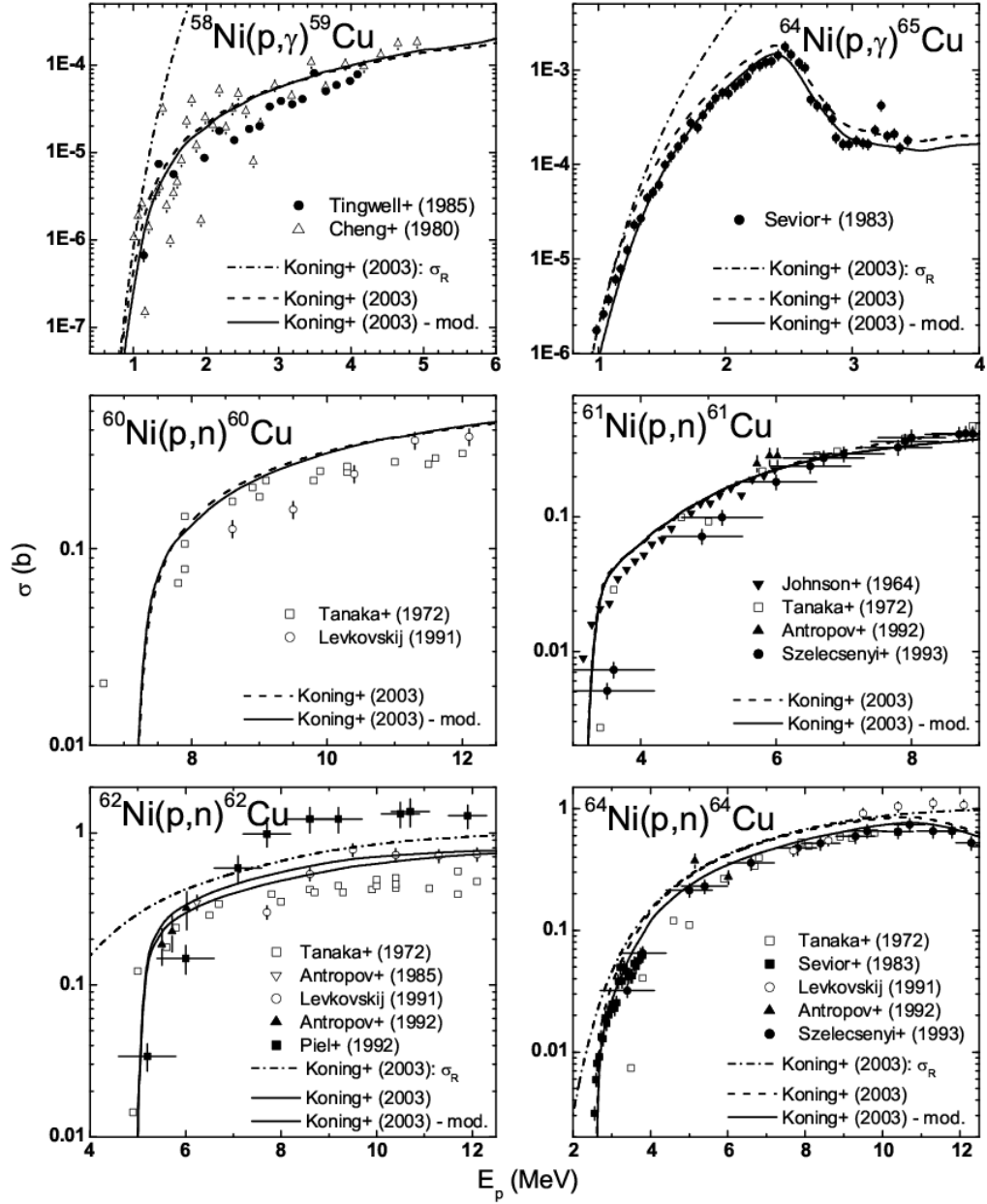,height=18.cm}}
\caption{As for Fig. 3, but for the Ni isotopes.}
\end{figure}

\newpage
\begin{figure}[t]	
\centerline{\epsfig{file=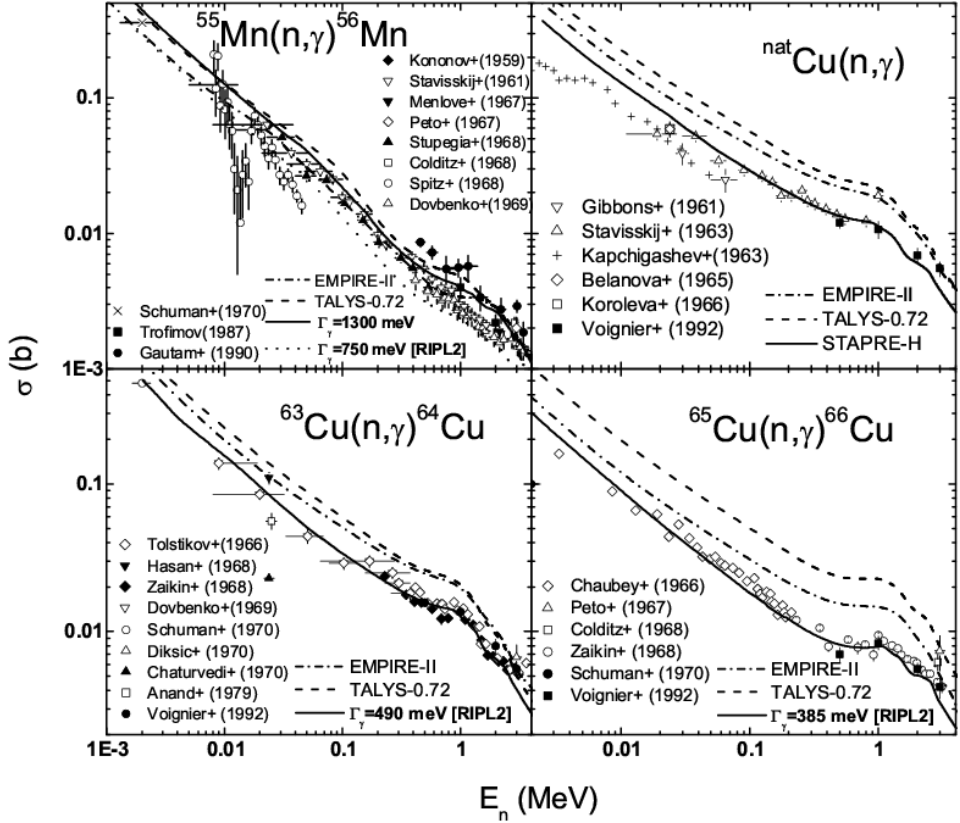,height=12.5cm}}
\caption{Comparison of the measured \cite{exfor} neutron-capture  
cross sections of $^{55}$Mn and $^{63,65,nat}$Cu target nuclei,
for incident energies up to 3--4 MeV, and calculated values by using 
the computer codes TALYS-0.72 (dashed curves) and EMPIRE-II 
(dash-dotted curves) with default global parameters, and the local
analysis with $\gamma$-ray strength functions $f_{E1}(E_{\gamma}$) 
within the EDBW model corresponding to either the experimental 
\cite{ripl2} average $s$-wave radiative widths $\Gamma_{\gamma0}$ 
(dotted curves), or $\Gamma_{\gamma0}$ values corresponding to a fit 
of experimental neutron capture data (solid curves).}
\end{figure}

\newpage
\begin{figure}[t]	
\centerline{\epsfig{file=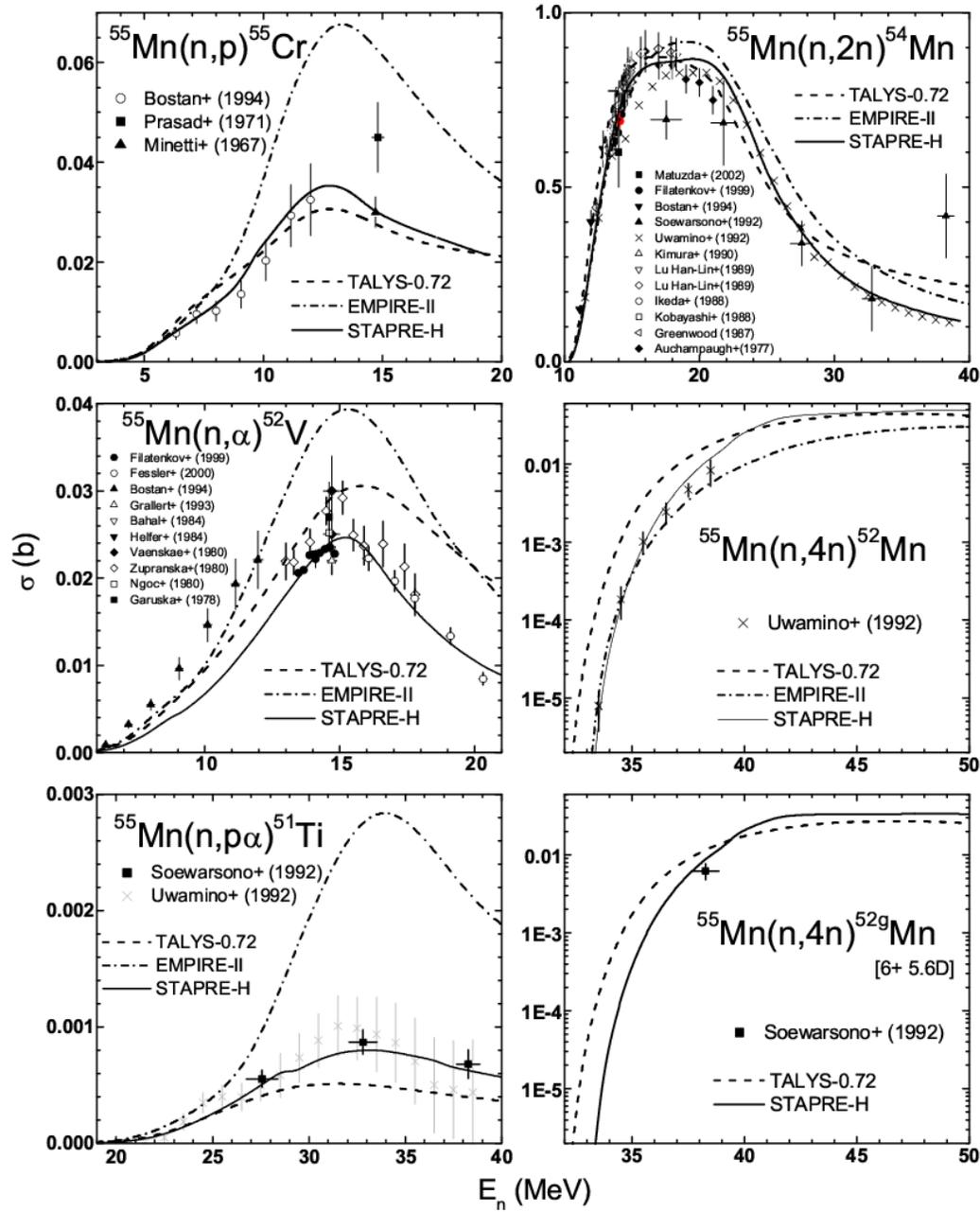,height=19.cm}}
\caption{Comparison of measured \cite{exfor} and calculated  
neutron-activation cross sections for the target nucleus $^{55}$Mn, 
by using the computer codes TALYS-0.72 (dashed curves) and EMPIRE-II 
(dash-dotted curves) with default global parameters, and STAPRE-H 
(solid curves) with the local parameter set given in this work.}
\end{figure}

\newpage
\begin{figure}[t]	
\centerline{\epsfig{file=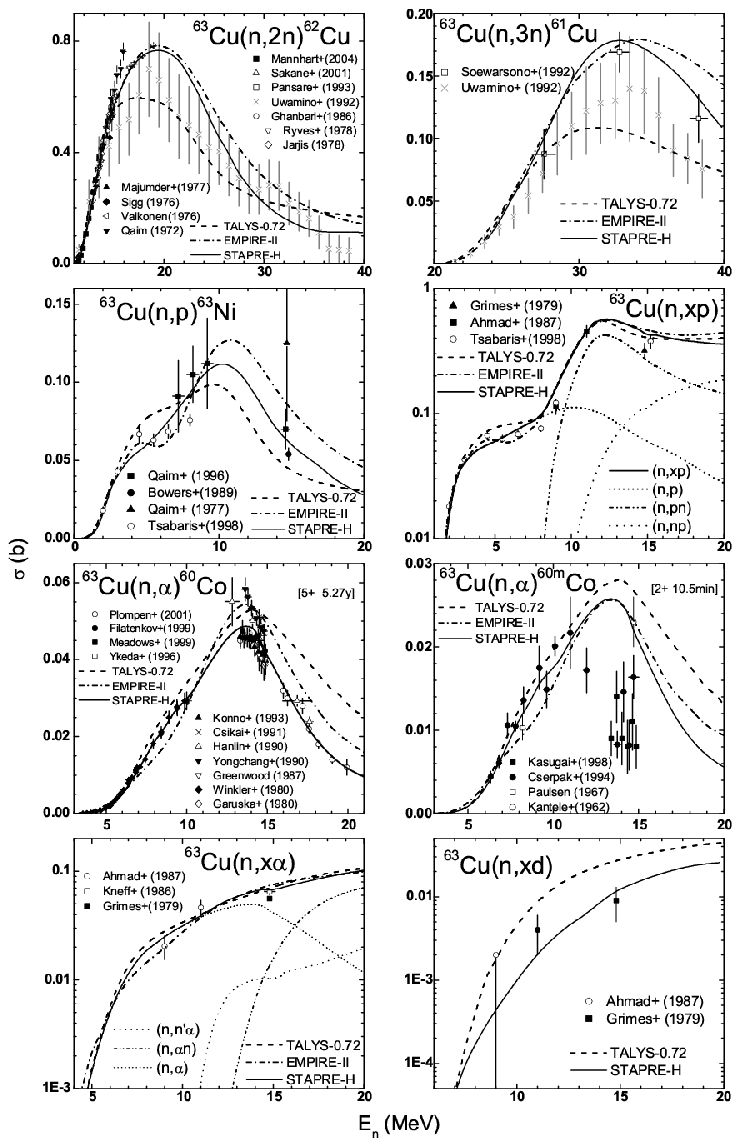,height=21.cm}}
\caption{As for Fig. 6, but for the target nucleus $^{63}$Cu.}
\end{figure}

\newpage
\begin{figure}[t]	
\centerline{\epsfig{file=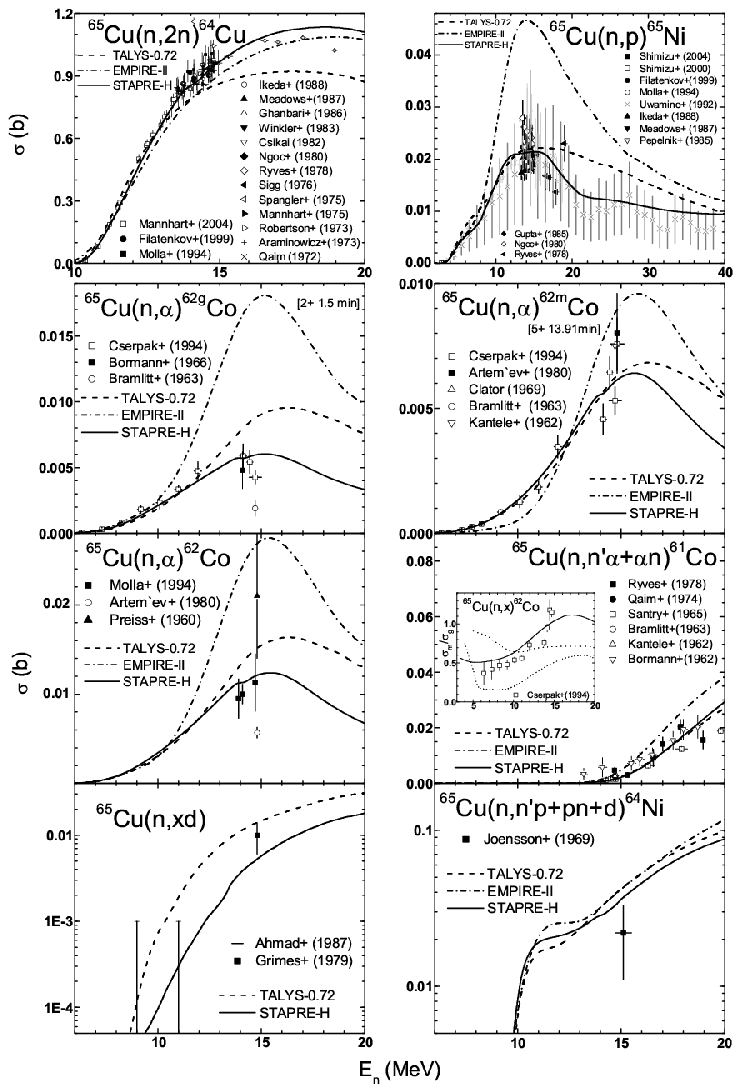,height=20.cm}}
\caption{As for Fig. 6, but for the target nucleus $^{65}$Cu.}
\end{figure}

\newpage
\begin{figure}[t]	
\centerline{\epsfig{file=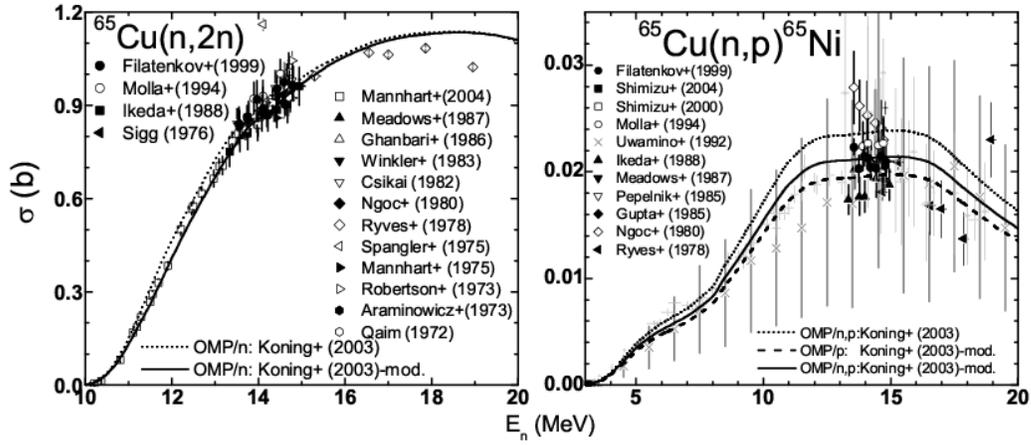,height=7.5cm}}
\caption{Comparison of measured \cite{exfor} and calculated cross 
sections within the local approach for the (n,2n) and (n,p) reactions 
on the target nucleus $^{65}$Cu, by using the OMP parameter sets of 
Koning and Delaroche \cite{ajk03} (dotted curves), and corresponding 
changes for the proton OMP (dashed curve) and neutron OMP (solid 
curves).}
\end{figure}

\newpage
\begin{figure}[t]	
\centerline{\epsfig{file=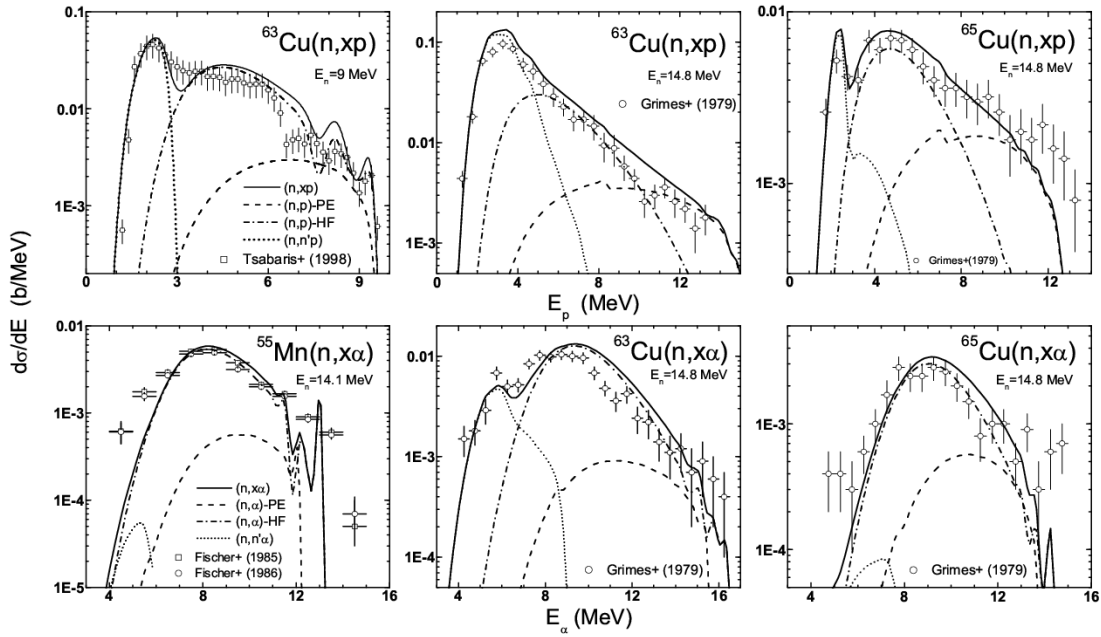,height=10.cm}}
\caption{Comparison of measured \cite{exfor,smg79,ct98,rf85,rf86} 
angle-integrated proton and $\alpha$-particle emission spectra from 
9, 14.1 and 14.8 MeV neutron-induced reactions on the $^{55}$Mn and 
$^{63,65}$Cu nuclei and calculated values within the local approach 
for the PE emission (dashed curves), statistical first- (dash-dotted 
curves) and second-emitted particles (dotted curves) from equilibrated 
compound nuclei, and their sum (solid curves).}
\end{figure}

\newpage
\begin{figure}[t]	
\centerline{\epsfig{file=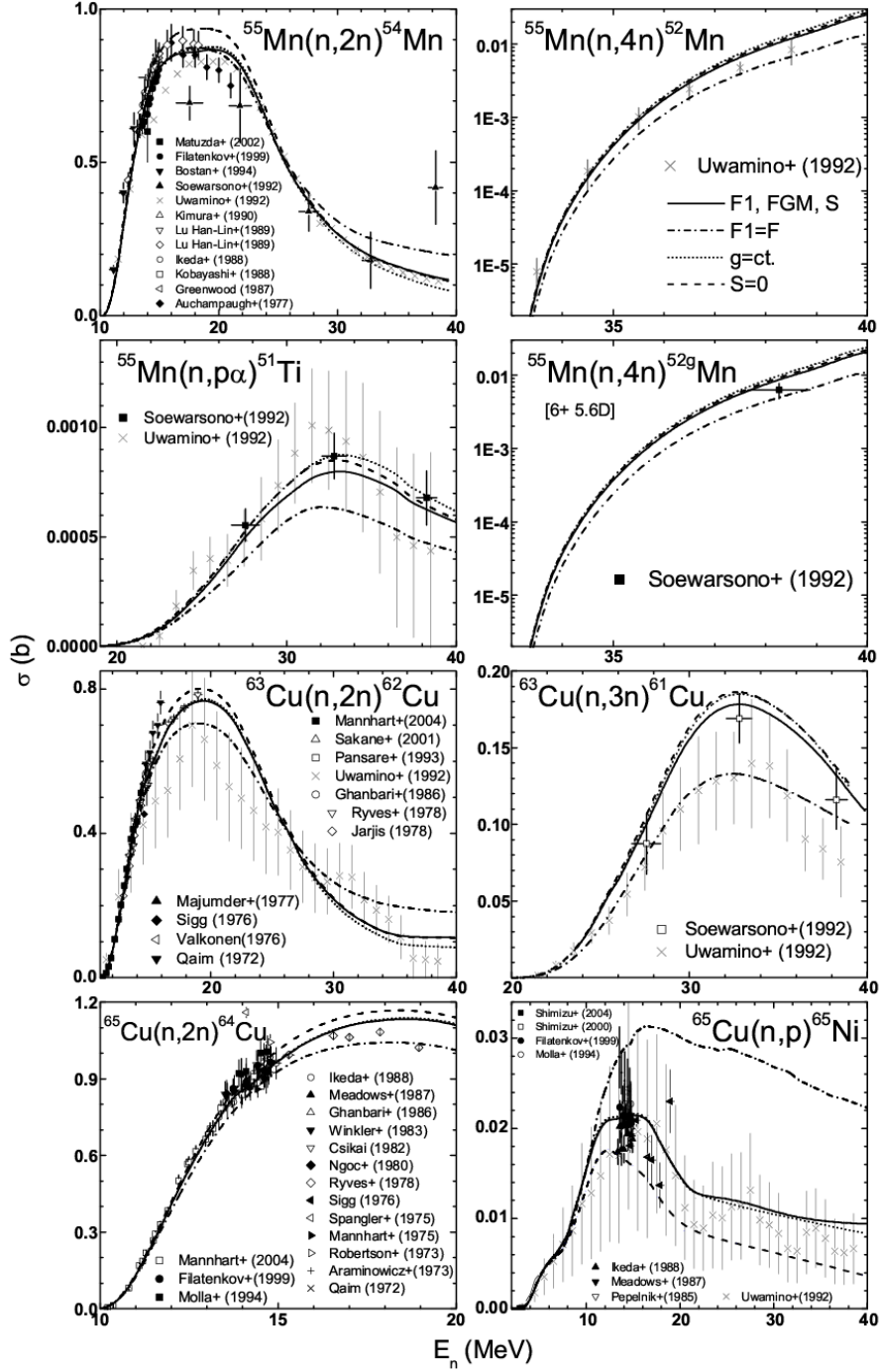,height=20.cm}}
\caption{Comparison of measured \cite{exfor} neutron-activation cross 
sections for the target nuclei $^{55}$Mn and $^{63,65}$Cu up to 40 MeV, 
and calculated values with the local parameter set given in this work
(solid curves) except for replacement of either the local-density Fermi 
energies $F_1(R_l)$ with the Fermi energy central value $F$ (dash-dotted
curves), or the average energy-dependent s.p.l. densities with the 
constant value $g(F)$ (dotted curves), as well as for removal of the 
shell correction $S$ in the PSD composite formula (dashed curves).}
\end{figure}

\newpage
\begin{figure}[t]	
\centerline{\epsfig{file=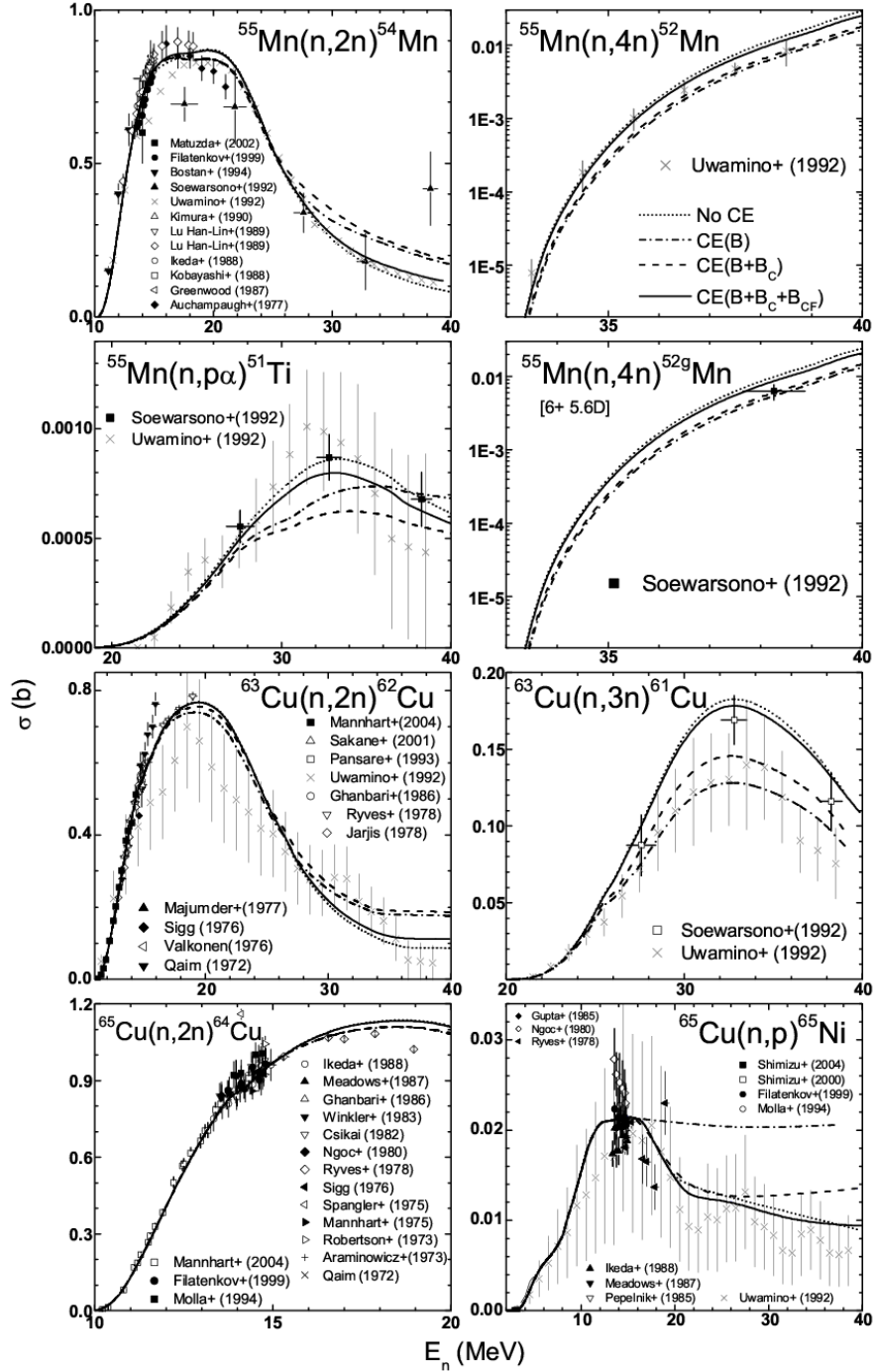,height=20.cm}}
\caption{As for Fig. 11, but for removal of the {\it 'continuum effect'} 
(CE) of the s.p.l. density within the particle-hole state density 
calculation (dotted curves), and taking into account for this effect the 
nucleon binding energy $B$ either alone (dash-dotted curves) or together 
with the Coulomb barrier $B_C$ (dashed curves), while the solid curves 
correspond to consideration of also the centrifugal barrier $B_{CF}$.}
\end{figure}

\end{document}